\documentclass[runningheads]{llncs}
\usepackage{graphicx}
\usepackage{xcolor}
\usepackage{xspace}
\usepackage{hyperref}
\usepackage{subcaption}
\usepackage{listings}
\usepackage{ulem}
\usepackage[utf8]{inputenc}

\hypersetup{
    colorlinks = true,
    linkcolor=red,
    citecolor=red,
    urlcolor=blue,
    linktocpage 
}

\lstdefinestyle{customC}{
  belowcaptionskip=1\baselineskip,
  breaklines=true,
  xleftmargin=\parindent,
  language=C,
  showstringspaces=false,
  basicstyle=\scriptsize\ttfamily,
  keywordstyle=\bfseries\color[rgb]{0.580, 0.000, 0.827},
  commentstyle=\textit{\color{green!40!black}},
  identifierstyle=\bfseries\color{cyan!75!black},
  stringstyle=\color{orange},
  deletekeywords={double,float},
  classoffset=1, 
  otherkeywords={double,float},
  morekeywords={double,float},
  keywordstyle=\bfseries\color{green!55!black},
  classoffset=0
}

\begin{document}


\title{Accurate simulation of operating system updates in neuroimaging using Monte-Carlo arithmetic}

\author{Ali Salari$^1$, Yohan Chatelain$^1$, Gregory Kiar$^2$, Tristan Glatard$^1$}

\authorrunning{A. Salari et al.}

\institute{$^1$Department of Computer Science and Software Engineering, Concordia University, Montréal, QC, Canada \\
$^2$Center for the Developing Brain, Child Mind Institute, New York, NY, USA \\
\email{m\_alari@encs.concordia.ca}}

\maketitle              

\begin{abstract}

  Operating system (OS) updates introduce numerical perturbations that impact
  the reproducibility of computational pipelines. In neuroimaging, this
  has important practical implications on the validity of computational
  results, particularly when obtained in systems such as high-performance
  computing clusters where the experimenter does not control software
  updates. We present a framework to reproduce the variability induced by OS
  updates in controlled conditions. We hypothesize that OS updates impact
  computational pipelines mainly through numerical perturbations originating
  in mathematical libraries, which we simulate using Monte-Carlo
  arithmetic in a framework called ``fuzzy libmath" (FL). We applied this methodology to pre-processing pipelines of
  the Human Connectome Project, a flagship open-data project in neuroimaging.
  We found that FL-perturbed pipelines accurately reproduce the
  variability induced by OS updates and that this similarity is only
  mildly dependent on simulation parameters. Importantly, we also found
  between-subject differences were preserved in both cases, though the
  between-run variability was of comparable magnitude for both FL and OS
  perturbations. We found the numerical precision in the HCP pre-processed
  images to be relatively low, with less than $8$ significant bits among the $24$
  available, which motivates further investigation of the numerical stability
  of components in the tested pipeline. Overall, our results establish that FL accurately simulates results
  variability due to OS updates, and is a practical framework to quantify
  numerical uncertainty in neuroimaging.

\end{abstract}

\section{Introduction}

Numerical round-off and cancellation errors are ubiquitous in
floating-point computations. In neuroimaging, they contribute to results
uncertainty along with other sources of variability, including population
selection, scanning devices, sequence parameters, acquisition noise, and
methodological
flexibility~\cite{bowring2019exploring,botvinik2020variability}. Numerical
errors manifest particularly through variations in elementary mathematical
libraries resulting from operating system (OS) updates. Indeed, due to
implementation differences, mathematical functions available in different
OS versions provide slightly different results. The impact of such
epsilonesque differences on image analysis depends on the conditioning of
the problem and the pipeline's numerical implementation. In neuroimaging,
established image processing pipelines have been shown to be substantially
impacted: for instance, differences in cortical thicknesses measured by the
same Freesurfer version in different execution platforms were shown to
reach statistical significance in some brain
regions~\cite{Gronenschild2012}, and Dice coefficients as low as 0.6 were
observed between FSL or Freesurfer segmentations obtained in different
platforms~\cite{Glatard2015,salari2020spot}. Such observations 
threaten the validity of neuroimaging results by revealing systematic
instabilities.

Despite its possible implications on results validity, the effect of OS
updates remains seldom studied due to (1) the lack of closed-form
expressions of condition numbers for complex pipelines and
non-differentiable non-linear analyses, (2) the technical challenge
associated with experimental studies involving multiple OS distributions
and versions, (3) the uncontrolled nature of OS updates.  As a result, the
effect of OS updates on neuroimaging analyses is generally neglected or
handled through the use of software containers (Docker or Singularity),
static executable builds, or similar approaches. While such techniques improve experiment portability,
they only mask numerical instabilities and do not tackle them. Numerical perturbations are bound to reappear
due to security updates~\cite{kaur2021analysis}, obsoleting software~\cite{perkel2020challenge}, or parallelization.
Therefore, the mechanisms through which numerical instabilities propagate need to be investigated and eventually addressed.

This paper presents ``fuzzy libmath" (FL), a framework to simulate OS updates in controlled conditions,
allowing software developers to evaluate the robustness of their tools with
respect to likely-to-occur numerical perturbations. As we hypothesize that
numerical perturbations resulting from OS updates primarily come from
implementation differences in elementary mathematical
libraries, we leverage Monte-Carlo arithmetic (MCA)~\cite{Parker1997-qq} to introduce
controlled amounts of noise in these libraries. FL
enables MCA in mathematical functions used by existing pipelines without the need to modify or recompile them.
To demonstrate the approach, we study the effect of
common OS updates on the numerical precision of structural MRI pre-processing
pipelines of the Human Connectome Project~\cite{van2013wu}, a major neuroimaging
initiative.

\section{Simulating OS updates with Monte-Carlo arithmetic}
\label{sec:MCA}

MCA models floating-point roundoff and cancellations errors through random
perturbations, allowing for the estimation of error distributions from
independent random result samples. MCA simulates computations at a given
\emph{virtual precision} using the following perturbation:
\begin{equation} \label{eq:mca_inexact}
  inexact(x) = x + 2^{e_x-t}\xi
  \label{mcadefinition}
\end{equation}
where $e_x$ is the exponent in the floating-point representation of $x$,
$t$ is the virtual precision and $\xi$ is a random uniform variable of
$(-\frac{1}{2}, \frac{1}{2})$.

MCA allows for three perturbation modes: Random Rounding (RR) introduces the
perturbation in function outputs, simulating roundoff errors; Precision Bounding
(PB) introduces the perturbation in function operands, allowing for the
detection of catastrophic cancellations; and, Full MCA combines RR and PB,
resulting in the following perturbation:
\begin{equation} \label{eq:mca_modes}
  mca\_mode(x \circ y) = inexact_{RR}(  inexact_{PB}(x) \circ inexact_{PB}(y) )
\end{equation}

To simulate OS updates, we introduce random perturbations in the GNU
mathematical library, the main mathematical library in GNU/Linux systems.
Instrumenting mathematical libraries with MCA raises a number of issues as
many functions assume deterministic arithmetic. For instance, applying random
perturbations around a discontinuity or within piecewise approximations
results in large variations and a total loss of significance that are not
relevant in our context. Therefore, we have applied MCA to proxy
mathematical functions wrapping those in the original library, such that only
the outputs of the original functions were perturbed but not their inputs or
the implementations themselves. This technique allows us to control the
magnitude of the perturbation as perceived by the application.

We instrumented the GNU mathematical library with MCA using
Verificarlo~\cite{denis2015verificarlo}, a tool that (1) uses the Clang
compiler to generate an LLVM (\url{http://llvm.org}) Intermediate
Representation (IR) of the source code, (2) replaces floating-point operations
in the IR by a call to the Verificarlo API, and (3) compiles the modified IR to
an executable using LLVM. The perturbation applied by the Verificarlo API
can be configured at runtime, for instance to change the virtual
precision applied to single- and double-precision floating-point values.

The resulting MCA-instrumented mathematical library, ``fuzzy libmath" (FL), is
loaded in the pipeline using \texttt{LD\_PRELOAD}, a Linux mechanism to
force-load a shared library into an executable. As a result, functions defined in fuzzy libmath
transparently overload the original ones without the need to modify
or recompile the analysis pipeline. Fuzzy libmath functions call the original
functions through \texttt{dlsym}, a function that returns the memory address of
a symbol. To trigger MCA instrumentation, a floating-point zero is added to the
output of the original function and the result of this sum is perturbed and
returned.

Finally, we measure results precision as the number of significant bits among result samples, as
defined in~\cite{Parker1997-qq}:
\begin{equation}
  s = -\log_2\left|\frac{\sigma}{\mu}\right|
  \label{estimatesig}
\end{equation}
where $\sigma$ and $\mu$ are the observed cross-sample standard deviation and average.

\section{HCP Pipelines \& Dataset}

\begin{figure}[t]
  \centering
  \includegraphics[width=\columnwidth]{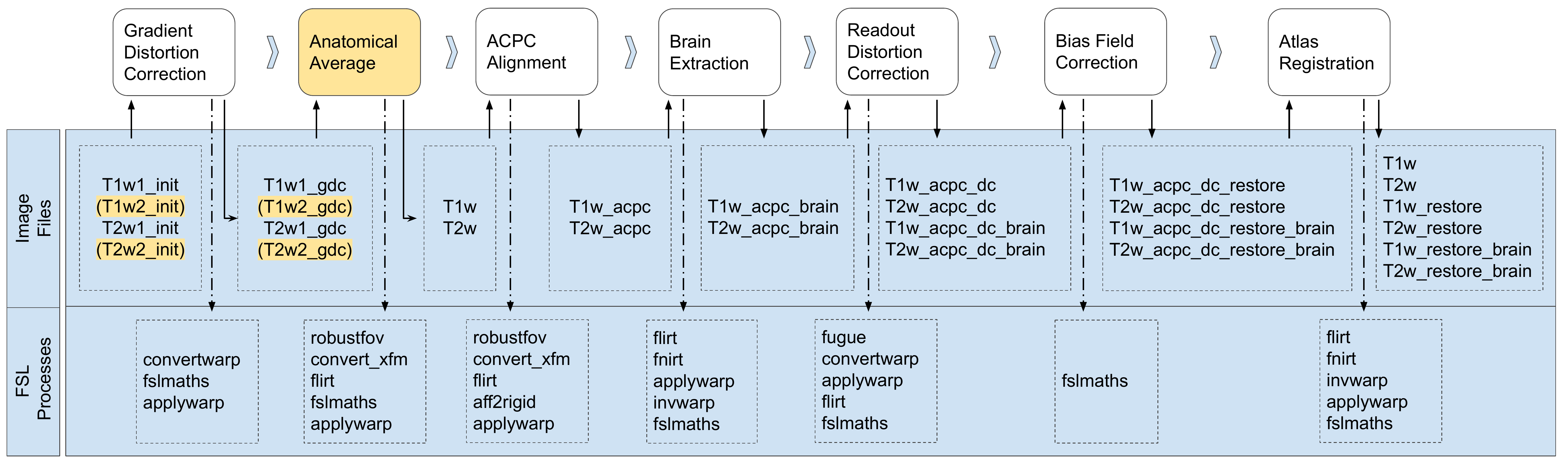}
  \caption{PreFreeSurfer pipeline steps.
  }
  \label{fig:pfs-steps}
\end{figure}

We apply the methodology described above to the minimal structural
pre-processing pipeline associated with the Human Connectome Project (HCP)
dataset~\cite{glasser2013}, entitled ``PreFreeSurfer''. This pipeline
consists of many independent components, including: spatial distortion
correction, brain extraction, cross-modal registration, and alignment to standard space. Each high-level component of this pipeline
(Fig.~\ref{fig:pfs-steps}) consists of several function calls using FSL, the FMRIB Software Library~\cite{jenkinson2012fsl}.
 The pipeline requires T1w and T2w images for each subject. A
full description of the pipeline is available
at~\cite{glasser2013}.

It should be noted that the PreFreeSurfer pipeline uses both single and double
precision functions from the GNU mathematical library. Among the pre-processing steps in the pipeline, it has been shown
that linear and non-linear registrations implemented in FSL
FLIRT~\cite{jenkinson2001global,jenkinson2002improved} and FNIRT~\cite{andersson2007non}
are the most sensitive to numerical instabilities~\cite{salari2020spot}. 

We selected $20$ unprocessed subjects from the HCP data release S500 available
in \href{https://db.humanconnectome.org}{the ConnectomDB repository}.
We selected these subjects from different subject types to cover execution paths sufficiently.
For each, the available data consisted of $1$ or $2$ T1w and T2w
images each, with spatial dimensions of $256 \times 320 \times 320$ and voxel
resolution of $0.7$ mm. Acquisition protocols and parameters are detailed
in~\cite{van2013wu}.
Two distinct experimental configurations were tested:
\begin{description}
  \item[Operating Systems (OS):] 
        subjects were processed on three different Linux
        operating systems inside Docker images: CentOS7 (glibc v.2.17), CentOS8 (glibc
        v.2.28), and Ubuntu20 (glibc v.2.31).
  \item[Fuzzy libmath (FL):] the dataset was processed on an Ubuntu20
        system using fuzzy libmath. The virtual precision ($t$) for the
        perturbations was swept from $53$ bits (the full mantissa for double-precision
        data) down to $1$ bit by steps of $2$. For $t>=24$ bits, only double-precision
        was altered and single-precision was set to 24 bits, and for $t<24$ bits,
        both double- and single-precision simultaneously were changed.
        Three FL-perturbed samples were generated for each subject and
        virtual precision, to match the number of OS samples.
\end{description}

After conducting both experiments, we selected the virtual precision that most
closely simulated the variability observed across OSes via the root-mean-square
error (RMSE) between the number of significant bits per voxel in all subjects
and conditions. This precision is referred to as the global nearest virtual precision
and was used to compare results obtained in both the FL and OS versions.

\section{Results}

The fuzzy libmath source code, Docker image specifications, and analysis code to
reproduce the results are available at \url{https://github.com/big-data-lab-team/MCA-libmath-paper}.
All experiments were conducted on the Béluga HPC computing cluster made
available by \href{https://www.computecanada.ca}{Compute Canada} through
\href{https://www.calculquebec.ca}{Calcul Québec}. Béluga is a general-purpose
cluster with $872$ available nodes. All nodes contain $2 \times$ Intel Gold 6148
Skylake @ 2.4 GHz ($40$ cores/node) CPU, and node memory can range between $92$
to $752$ GB. The average processing time of the pipeline without FL
instrumentation was $69$ minutes (average of 3 executions). The FL perturbation increased it to $93$ minutes.

We ensured that the pipeline does not use pseudo-random numbers by processing each subject twice on the same operating system.
To validate that FL was correctly instrumented with
Verificarlo, we used Veritracer~\cite{chatelain2018veritracer}, a tool for
tracing the numerical quality of variables over time.
For one subject, the traces showed that the number of significant bits
in the function outputs varied over time, confirming the instrumentation with MCA.
Throughout the pipeline execution, Veritracer reported approximately $4$~billion calls to FL,
with the following ratio of calls: $47.12\%$ \texttt{log},
$40.96\%$ \texttt{exp}, $6.92\%$ \texttt{expf}, $3.39\%$ \texttt{logf},
$1.55\%$ \texttt{sincosf}, and $0.06\%$ of cumulated calls to \texttt{atan2f},
\texttt{pow}, \texttt{sqrt}, \texttt{exp2f}, \texttt{powf}, \texttt{log10f}, \texttt{log10}, \texttt{cos}, and \texttt{asin}.
We also checked that long double types were not used.

\subsection{Fuzzy libmath accurately simulates the effect of OS updates}
\begin{figure}[b]
  \begin{subfigure}[t]{0.52\linewidth}
    \centering
    \includegraphics[width=\linewidth]{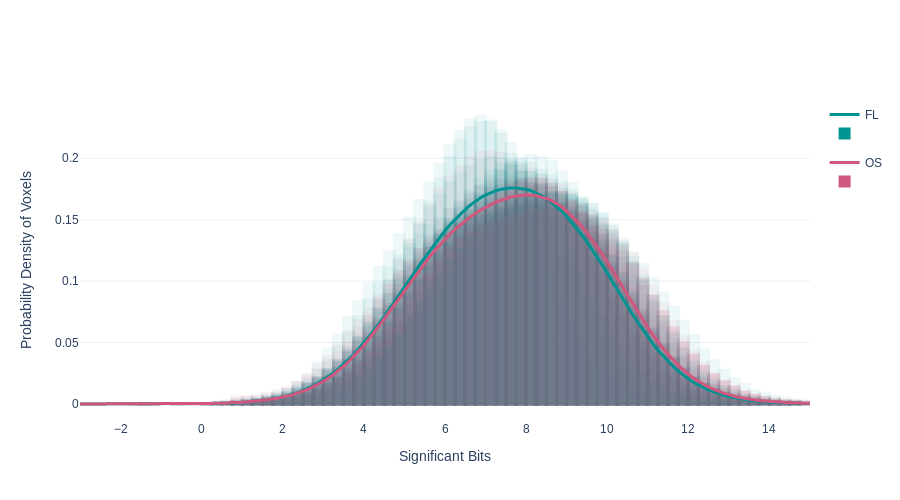}
    \caption{Distribution of significant bits}
    \label{fig:sigbits}
  \end{subfigure}
  \hfill
  \begin{subfigure}[t]{0.45\linewidth}
    \centering
    \includegraphics[width=\linewidth]{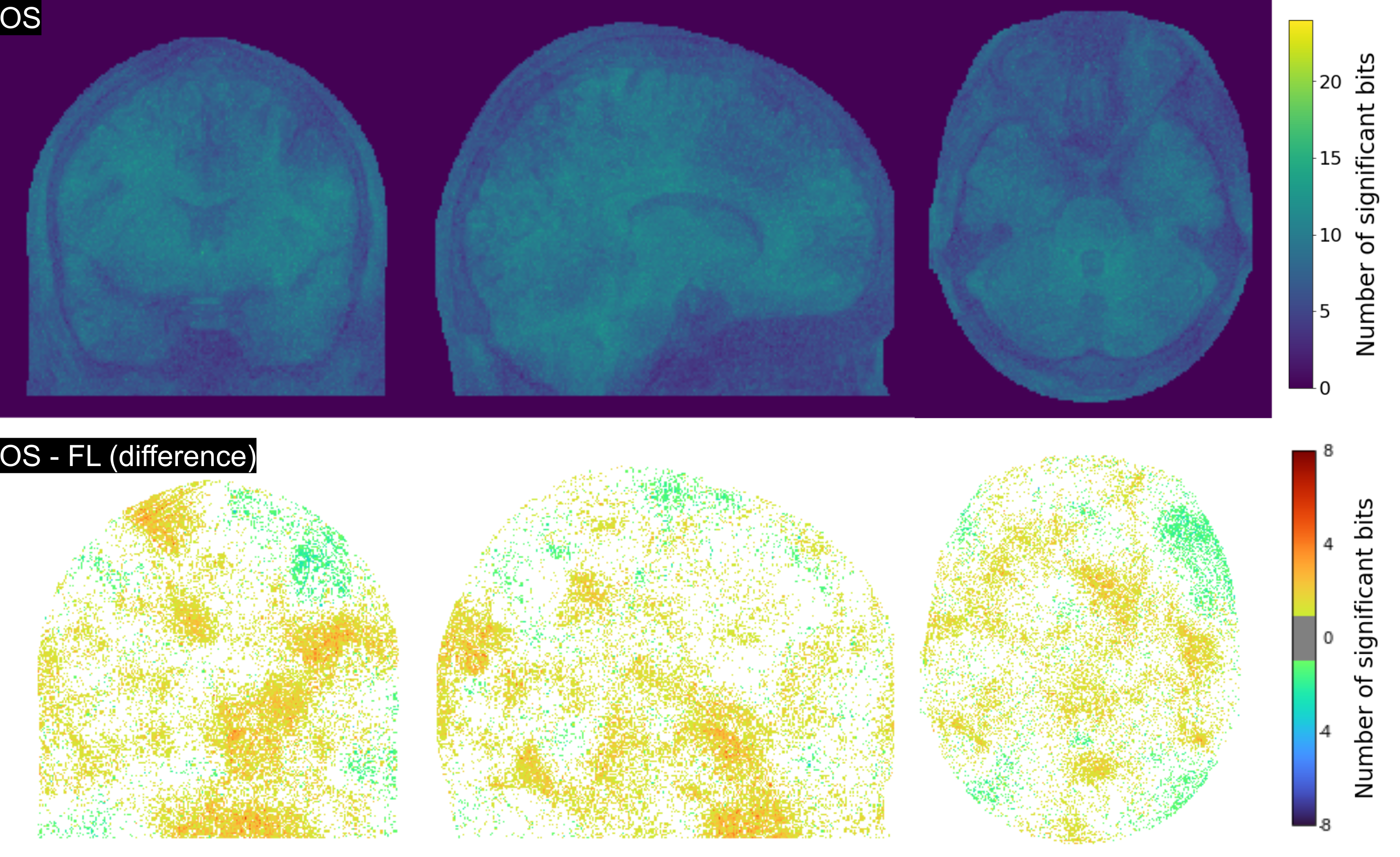}
    \caption{Significance map (subject average)}
    \label{fig:mnispace}
  \end{subfigure}
  \caption{Comparison of OS and FL effects on the precision of PreFreeSurfer results
   for n=20 subjects.
    FL samples were obtained at the global nearest virtual precision of t=37 bits. }
  \label{fig:mca-os-comp}
\end{figure}
Fuzzy libmath accurately reproduced the effect of OS updates, both globally
(Fig.~\ref{fig:sigbits}) and locally (Fig.~\ref{fig:mnispace}). The
distributions of significant bits in the atlas registered T1w images were nearly
identical ($p > 0.05$, KS test) on the average and individual subject distributions for $15/20$ subjects,
after correcting for multiple comparisons. Locally, the
spatial distribution of significant digits also appeared to be preserved. Losses
in significance were observed mainly at the brain-skull interface and between
brain lobes, indicating spatial dependency of numerical properties.

The average number of significant bits in either the FL or OS conditions
were $7.76$ out of $24$ available, which corresponds to $2.32$ significant (base $10$) digits.
This relatively low precision motivates future investigations of the
stability of pipeline components, in particular for image registration.

\subsection{Fuzzy libmath preserves between-subjects image similarity}

\begin{figure}
  \centering
  \includegraphics[width=\columnwidth]{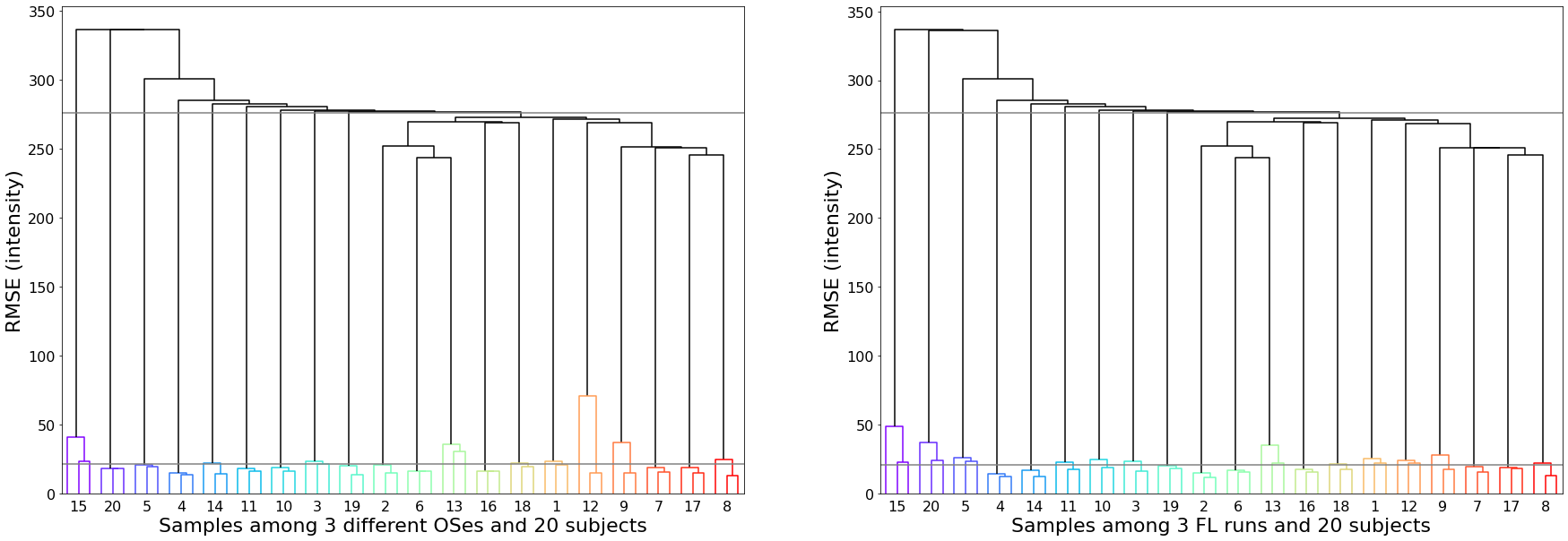}
  \caption{RMSE-based hierarchical clustering of OS (left) and FL
    (right) samples. Colors identify different subjects, showing that
    similarities between subjects are preserved by the numerical perturbations. Horizontal gray lines
    represent average RMSEs between (top line) and within (bottom line) subject clusters.}
  \label{fig:clusters}
\end{figure}

Numerically-perturbed samples remained primarily clustered by individual subjects
(Fig.~\ref{fig:clusters}), indicating that neither FL nor OS perturbations were
impactful enough to blur the differences between subjects.  Notably, the similarity between subjects was
also preserved by the numerical perturbation, leading to the same subject
ordering in the dendrograms. However, the average
RMSE within samples of a given subject was approximately $13 \times$
lower than the average RMSE between different subjects. The fact that between-subject variabilities
were nearly on the same order of magnitude as OS and FL variability demonstrates
the potential severity of these instabilities.

\subsection{Results are stable across virtual precision}

\begin{figure}[t]
  \centering
  \includegraphics[width=.7\columnwidth]{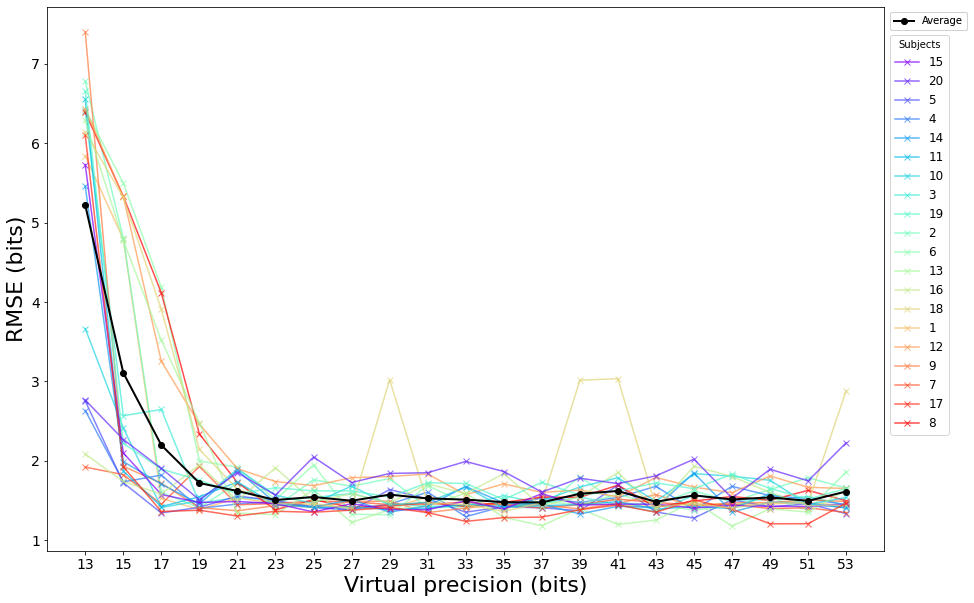}
  \caption{Comparison of RMSE values computed between OS and FL results for different virtual precisions.}  \label{fig:vprecision}
\end{figure}

The FL results presented previously were obtained at the global nearest virtual precision
of t=$37$~bits, determined as the precision which minimized the RMSE
between FL and OS average maps of significant bits. We varied the
virtual precision in steps of $2$ between t=$1$ and t=$53$~bits
(Fig.~\ref{fig:vprecision}). On average, no noticeable RMSE change was
observed between the FL and OS variability for precisions ranging from
t=$21$ to t=$53$~bits, which shows that FL can robustly approximate OS updates.

The observed plateau suggests the existence of an ``intrinsic precision" for the pipeline, above
which no improvement in results precision is expected. For the tested pipeline,
this intrinsic precision was observed at t=$21$ bits, which indicates that
the pipeline could be implemented exclusively with single-precision
floating-point representations ($24$ bits of mantissa) without loss of
results precision. This would substantially decrease the pipeline memory
footprint and computational time, as approximately $88\%$ of operations used in this pipeline
made use of double-precision data.
In addition, the presence of such a
plateau suggests that numerical perturbations introduced by OS updates
might be in the range of machine error (t=$53$~bits), although it is also
possible that the extent of the plateau results from the numerical conditioning of the
tested pipeline. It is possible in contrast that the absence of such a plateau
would suggest an unstable pipeline that would benefit either from correction or larger datatypes.
The ability to capture stability across a range of precisions importantly
demonstrates a key advantage of using FL to simulate OS variability. 

The relationship between RMSE of individual subjects was generally consistent with the
average line, with the notable exception of subject $18$. The observed discrepancies between
this subject and potential others might be leveraged for quality control checks and, as a
result, inform tool development.

The pipeline failed to complete for at least one subject below the virtual precision of t=$13$~bits, also
referred to as the tolerance of the pipeline.
Specifically, $51\%$ of pipeline executions crashed among all subjects for precisions ranging from $1$--$11$~bits,
and there was no relationship between tolerance-level and precision.
The error raised was in the Readout Distortion Correction portion of the pipeline, and appears to stem from the
FSL FAST tissue segmentation. The specific source of the error within this component is presently unknown, but
is an open question for further exploration.

\section{Conclusion \& Discussion}

We demonstrated fuzzy libmath as an accurate method to simulate variability in neuroimaging results due to OS
updates. Alongside this evaluation, fuzzy libmath can be
used by pipeline developers or consumers to evaluate the numerical uncertainty of tools and results.
 Such evaluations may also help decrease pipeline memory usage and computational time
through the controlled use of reduced numerical precision.
Fuzzy libmath does not require any modification of the pipeline
as it operates on the level of shared libraries. The accuracy of
the simulations were shown to be robust across a wide range of virtual
precisions, which reinforces the applicability of the method.

The proposed technique is directly applicable to MATLAB code executed with GNU Octave, 
to Python programs executed on Linux, and to C programs that depend on GNU libmath.
Numerical noise can be introduced in other libraries, such as OpenBLAS or NumPy, using our \url{https://github.com/verificarlo/fuzzy} environment.

A commonly used approach to address instabilities resulting from OS
version updates in practice is to sweep the issue under the rug of software containers
or static linking. While such solutions are undoubtedly helpful to improve
code portability or strict re-executability, a more honest position is to consider computational
results as realizations of random variables depending on numerical error.
The presented technique enables estimating result distributions, a first
step toward making analyses reproducible across heterogeneous execution
environments. While this work did not investigate the precise cause of
numerical instabilities by tracing the system function calls, this is a topic for future work.

The tested OS versions span a timeframe of 7 years
(2012--2020) and focused on GNU/Linux, a widely-used platform in neuroimaging~\cite{hanke2011neuroscience}.
Given that our experiments focused on numerical perturbations applied to mathematical functions, 
which are implemented similarly across OSes,
our findings are likely to generalize 
to OS/X or MS Windows, although future work would be needed to confirm that. 
The tested pipeline is the official solution of the HCP
project to pre-process data, and is considered the state-of-the-art. This pipeline
assembles software components from the FSL
toolbox consistent with common practice in neuroimaging, such as in
fMRIPrep~\cite{esteban2019fmriprep} or the FSL feat workflow~\cite{jenkinson2012fsl},
to which fuzzy libmath can be directly applied. 
 Efforts are on-going to use fuzzy libmath in fMRIPrep software tests,
 to guarantee that bug fixes do not perturb results beyond numerical uncertainty. 

The fact that the induced numerical variability preserves
image similarity between subjects is reassuring and, in fact, exciting. OS updates
provide a convenient, practical target to define a virtual precision
leading to a detectable but still reasonable numerical perturbation. However, it is also
of importance that OS- and FL-induced variability were on a similar
order of magnitude as subject-level effects. This suggests that the preservation of relative
between-subject differences may not hold in all pipelines, and such a comparison could be
used to evaluate the robustness of a pipeline to OS instabilities.
The fact that the results observed across OS versions and FL perturbations arise
from equally-valid numerical operations also suggests that the observed variability
may contain meaningful signal.
In particular, signal measured from these perturbations might be leveraged to
enhance biomarkers, as suggested in~\cite{kiar2020data} where augmenting a
diffusion MRI dataset with numerically-perturbed samples was shown to
improve age classification.

%
%
%
\bibliographystyle{splncs04}
\bibliography{biblio}

\end{document}